\documentclass{PoS}
\usepackage{slashed}
\usepackage[dvipsnames]{xcolor}
\usepackage{tikz}
\usetikzlibrary{arrows,calc,positioning}
\tikzset{
>=stealth',
help lines/.style={dashed, thick},
axis/.style={<->,thick},
important line/.style={thick},
connection/.style={thick, dotted},
every picture/.style={/utils/exec={\sffamily}},
}

\title{
Meson correlation functions at high temperature QCD:
$SU(2)_{CS}$ symmetry vs. free quarks
}

\ShortTitle{
Meson correlation functions at high temperature QCD
}

\author{\speaker{C. Rohrhofer}$^a$,
        Y.~Aoki$^b$$^c$,
        G.~Cossu$^d$,
        L.Ya.~Glozman$^a$,
        S.~Hashimoto$^b$$^e$,
        S.~Prelovsek$^f$$^g$$^h$,\\
        $^a$ Institute of Physics, University of Graz, 8010 Graz, Austria\\
        $^b$ KEK Theory Center, High Energy Accelerator Research Organization (KEK), Tsukuba 305-0801, Japan\\
        $^c$ RIKEN BNL Research Center, Brookhaven National Laboratory, Upton NY 11973, USA\\
        $^d$ School of Physics and Astronomy, The University of Edinburgh, Edinburgh EH9 3JZ, United Kingdom\\
        $^e$ School of High Energy Accelerator Science, The Graduate University for Advanced Studies (Sokendai), Tsukuba 305-0801, Japan\\
        $^f$ Faculty of Mathematics and Physics, University of Ljubljana, 1000 Ljubljana, Slovenia\\
        $^g$ Jozef Stefan Institute, 1000 Ljubljana, Slovenia\\
        $^h$ Institute f\"ur Theoretische Physik, Universit\"at Regensburg, D-93040, Germany\\
        }

\abstract{
We report on the progress of understanding spatial correlation functions in
high temperature QCD. We study isovector meson operators in $n_f=2$ QCD using
domain-wall fermions on lattices of $N_s=32$ and different quark masses. It has
previously been found that at $\sim 2T_c$ these observables are not only
chirally symmetric but in addition approximately $SU(2)_{CS}$ and $SU(4)$
symmetric. In this study we increase the temperature up to $5T_c$ and can
identify convergence towards an asymptotically free scenario at very high
temperatures.
}

\FullConference{The 36th Annual International Symposium on Lattice Field Theory - LATTICE2018\\
		22-28 July, 2018\\
		Michigan State University, East Lansing, Michigan, USA.}

\begin{document}

\section{Introduction}
\label{intro}

In a series of numerical experiments, initiated in Ref.~\cite{Lang:2011vw},
some interesting findings have been made
\cite{Denissenya:2014poa,Denissenya:2014ywa,Denissenya:2015mqa,Denissenya:2015woa}:
Upon truncating the low-modes of the Dirac operator the
spectrum of $J=1,2$ mesons and light baryons revealed a symmetry larger
than the chiral symmetry of QCD, see Fig. 1 for the  $J=1$ results. Apriori one
expects that such a truncation would lead to a restoration of chiral and possibly of
$U(1)_A$ symmetries in hadrons, if hadrons survive this procedure. Surprisingly
larger symmetries, $SU(2)_{CS}$ and $SU(2n_f)$, emerge that contain chiral symmetries
of the QCD Lagrangian as subgroups \cite{Glozman:2014mka,Glozman:2015qva}. These symmetries
are symmetries of the chromo-electric interaction in QCD, while the chromo-magnetic
interaction as well as the quark kinetic term break them. From these results one can
conclude that the effects of the chromo-magnetic interaction in QCD are located
exclusively in the near-zero modes, while confining chromo-electric interaction is
distributed among all modes of the Dirac operator.
While chiral  symmetries for meson propagators have been
 analytically shown to arise upon truncation
of the low-lying modes, emergence of $SU(2)_{CS}$ and $SU(2n_f)$ is encoded
in some matrix elements that include some unknown microscopic dynamics \cite{Lang:2018vuu}.
Given this insight one could expect
emergence of the same symmetries in QCD at high temperature without any
truncation because the high temperature naturally suppresses the near-zero modes
of the Dirac operator.

Recently a full $n_f=2$ flavor simulation of QCD using chirally-symmetric Domain-wall fermions
at temperatures above the
chiral restoration temperature
has shown that the spatial correlators of isovector $J=1$ mesons approximately feature
$SU(2)_{CS}$ and $SU(2n_f)$ symmetries \cite{Rohrhofer:2017grg}.
In the present study we extend the temperature range up to $\sim 5.5T_c$.

\begin{figure}
\centering
\includegraphics[scale=0.65]{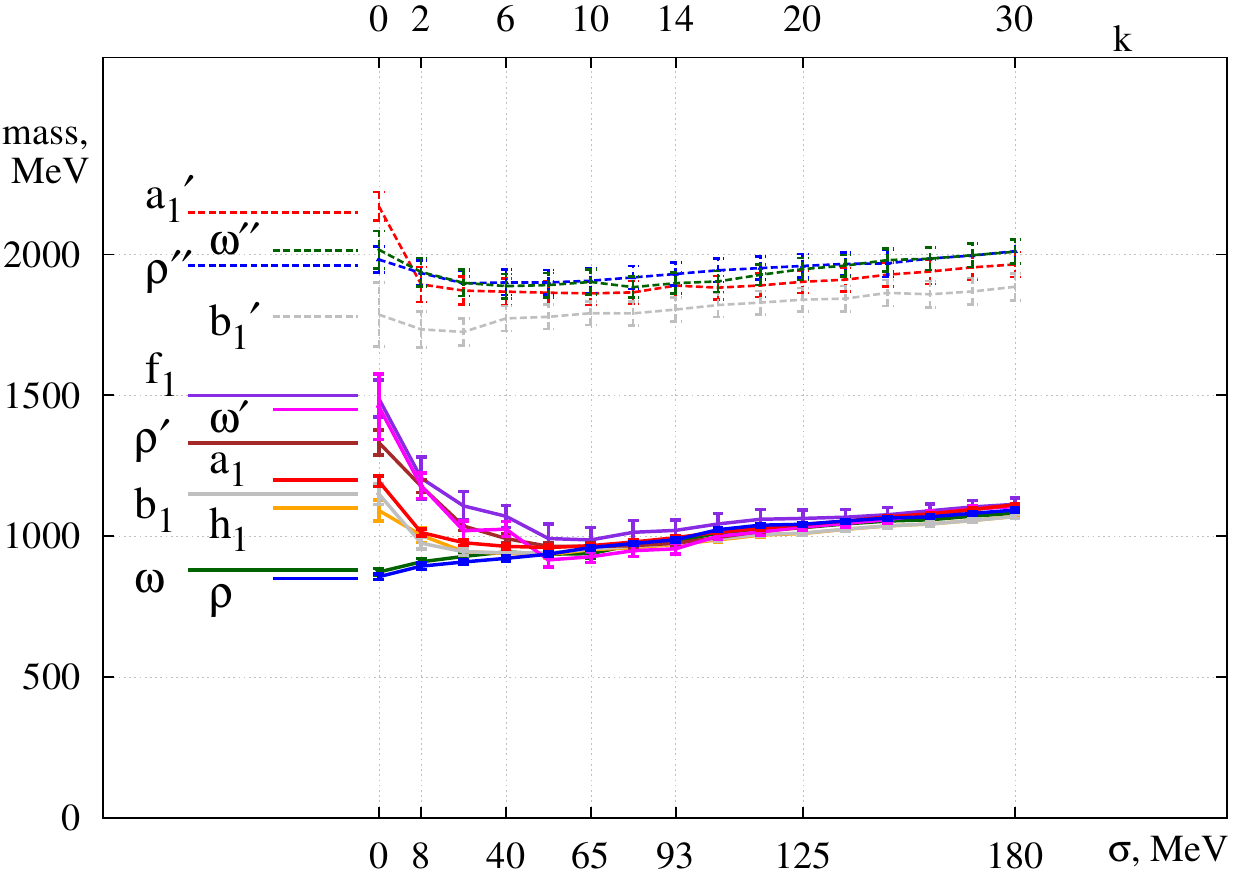}
\caption{$J=1$ meson spectrum upon low-mode truncation of the Dirac operator.
Figure from Ref. \cite{Denissenya:2014ywa}.}
\label{lowmodetruncation}
\end{figure}

\section{Method}
\label{method}
Our lattice setup consists of two mass-degenerate flavors of light quarks, which
use the M\"obius domain wall fermion discretization \cite{Brower:2005qw,Brower:2012vk}.
The gauge sector is simulated using the tree-level improved Symanzik action. The quark masses are
set to values between 2--15 MeV, and the temperature is set by varying strong coupling $\beta$ and $N_t$.
More details concerning this set of parameters can be found in Ref.~\cite{Tomiya:2016jwr,Cossu:2015kfa}
and Table~\ref{params}. The  (pseudo)critical temperature for this ensembles is $T_c=175\pm5$ MeV.  
In total this allows us to access temperatures from
$T \simeq 1.2-5.5 T_c$.

 \begin{table}
 \center
 \begin{tabular}{cc|c|c|c|c|cc}
 \hline\hline
 $N_s^3\times N_t$ & $\beta$ & $m_{ud}a$ & $a$ [fm] & \# configs & $L_s$ & $T$ [MeV] & $T/T_c$ \\\hline
 $32^3\times 8$ & $4.10$  & 0.001 & $0.113$ &     800  & 24  & $\sim220$ & $\sim 1.2$ \\
 $32^3\times 8$ & $4.18$  & 0.001 & $0.096$ &     230  & 12  & $\sim260$ & $\sim 1.5$ \\
 $32^3\times 8$ & $4.30$  & 0.001 & $0.075$ &     260  & 12  & $\sim320$ & $\sim 1.8$ \\
 $32^3\times 8$ & $4.37$  & 0.001 & $0.065$ &      77  & 12  & $\sim380$ & $\sim 2.2$ \\
 $32^3\times 6$ & $4.30$  & 0.001 & $0.075$ &     270  & 12  & $\sim440$ & $\sim 2.5$ \\
 $32^3\times 8$ & $4.50$  & 0.001 & $0.051$ &     197  & 12  & $\sim480$ & $\sim 2.7$ \\
 $32^3\times 4$ & $4.30$  & 0.001 & $0.075$ &     200  & 10  & $\sim660$ & $\sim 3.8$ \\
 $32^3\times 4$ & $4.50$  & 0.001 & $0.051$ &     209  & 10  & $\sim960$ & $\sim 5.5$ \\
 \hline\hline
 \end{tabular}
 \caption{Lattices used in this work:
 $L_s$ is the length in the fifth dimension of the M\"obius domain wall fermions.
 }
 \label{params}
 \end{table}

In this work we focus on the local isovector bilinears
\begin{equation}
\mathcal{O}_\Gamma(x) = \bar q(x) \vec\tau \; \Gamma q(x),
\end{equation}
where $\Gamma$ denotes any element of the Clifford algebra and
$\vec \tau$ are the generators in flavor space. For spatial measurements
in $z$-direction we sum over orthogonal lattice slices in $x, y, t$-direction
\begin{equation}
C_\Gamma ( n_z ) = \sum \limits_{n_x, n_y, n_t}
\langle
\mathcal{O}_\Gamma(n_x,n_y,n_z,n_t)
\mathcal{O}_\Gamma(\mathbf{0},0)^\dagger
\rangle.
\end{equation}
The gamma structures are grouped into the following objects:
\begin{itemize}
\item Scalar ($S$, $\Gamma=\mathbf{1}$) and Pseudoscalar (PS, $\Gamma=\gamma_5$)
\item Vector ($\mathbf{V}$, $\Gamma=\gamma_k$) and Axial-vector ($\mathbf{A}$, $\Gamma=\gamma_k\gamma_5$)
\item Tensor-vector ($\mathbf{T}$, $\Gamma=\gamma_k\gamma_3$) and Axial-tensor-vector ($\mathbf{X}$, $\Gamma=\gamma_k\gamma_3\gamma_5$)
\end{itemize}
as well as the non-propagating $\Gamma=\gamma_3$ and $\Gamma=\gamma_3 \gamma_5$.
The vector indices take on values $k=1,2,4$, which we denote as $x,y$ and $t$ respectively.
\textit{E.g.} the first component of the Axial-vector would be $A_x$.

\section{Results}
\label{results}
\begin{figure}
\centering
\includegraphics[scale=0.32]{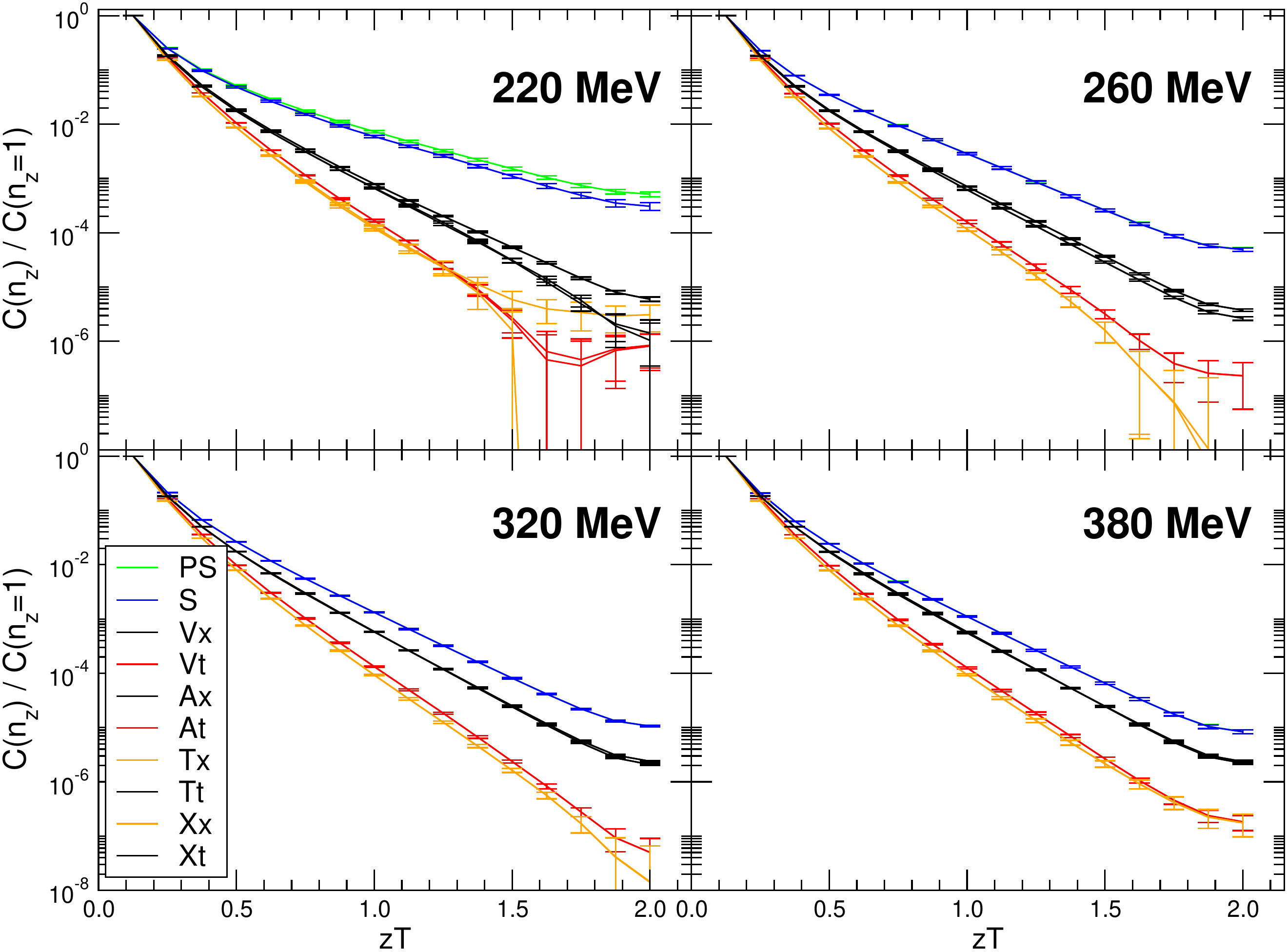}
\caption{Normalized spatial correlation functions for operators
and ensembles up to $\sim 380$ MeV.}
\label{corrs}
\end{figure}

\begin{figure}
\centering
\includegraphics[scale=0.33]{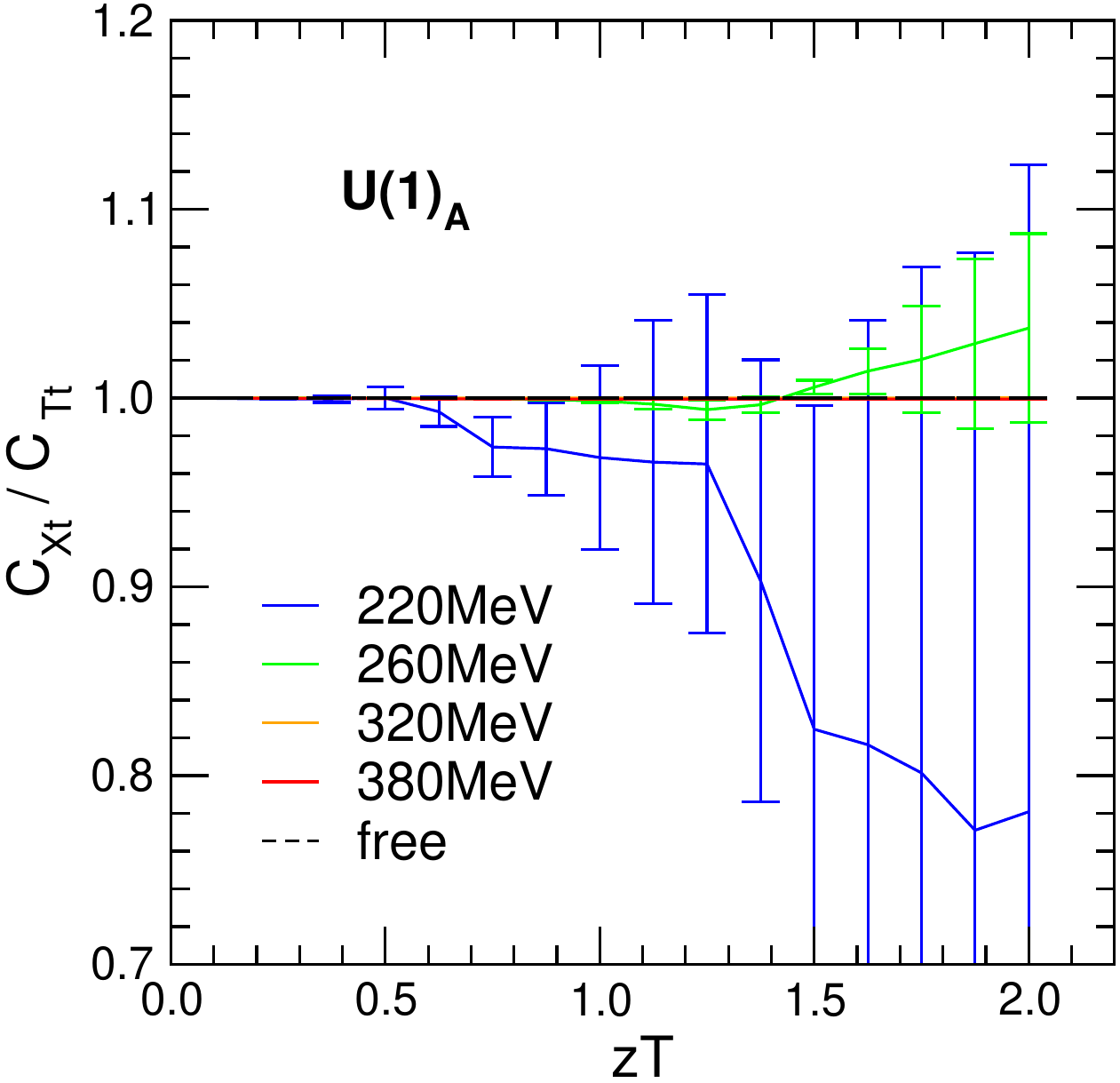}
\includegraphics[scale=0.33]{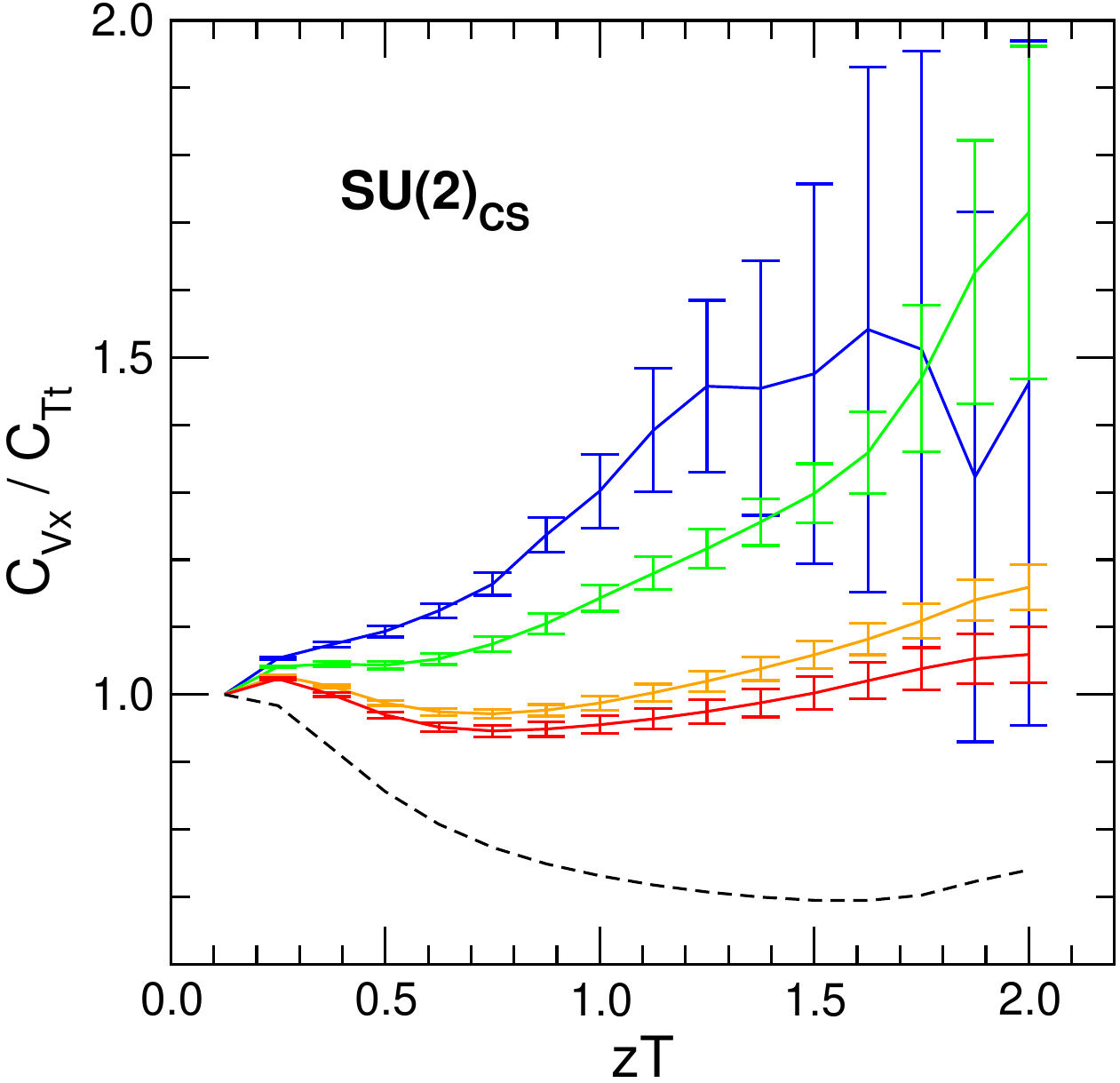}
\caption{
Detailed ratio for correlation functions of operators connected by
$U(1)_A$ as well as $SU(2)_{CS}$ symmetries.
}
\label{ratios}
\end{figure}

Figures \ref{corrs} and \ref{ratios}
summarize findings for $T \leq 2T_c$ from \cite{Rohrhofer:2017grg}.
Fig.~\ref{corrs} shows
that correlation functions of operators connected by both $U(1)_A$ transformations
($PS$ and $S$, $\mathbf{T}$ and $\mathbf{X}$)
as well as $SU(2)_L \times SU(2)_R$ symmetry
($\mathbf{V}$ and $\mathbf{A}$)
get degenerate at temperatures $T>220$ MeV.
This is a clear signal for restoration of the flavor chiral symmetry, as well as for 
at least approximate restoration of $U(1)_A$ symmetry.

However a larger degeneracy is seen. E.g. correlators for $V_x$ and $T_t$, as well as $V_t$ and $T_x$
get approximately degenerate at $T \sim 380$ MeV. Such degeneracies indicate emergence
of the $SU(2)_{CS}$ and $SU(4)$.
Fig.~\ref{ratios} shows the ratio of corresponding correlators connected by $U(1)_A$ and
$SU(2)_{CS}$ transformations. It can be seen that  while the $U(1)_A$ symmetry is "exactly"
restored,  a degeneracy of correlators connected by the $SU(2)_{CS}$ transformation is only approximate:
There are still symmetry breaking effects at the level of 5\%.

\begin{figure}
  \centering
  \includegraphics[scale=0.38]{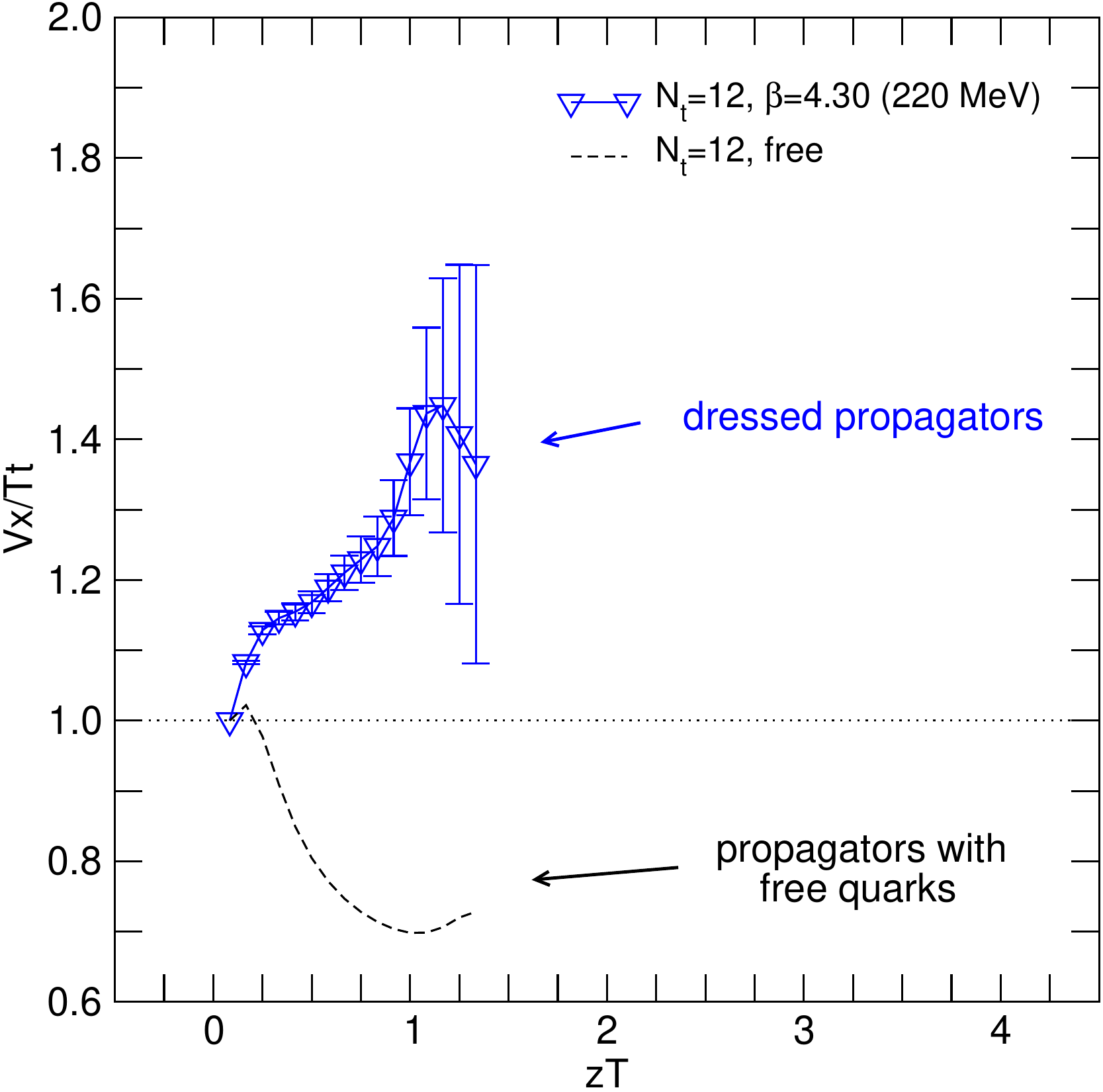}
  \includegraphics[scale=0.38]{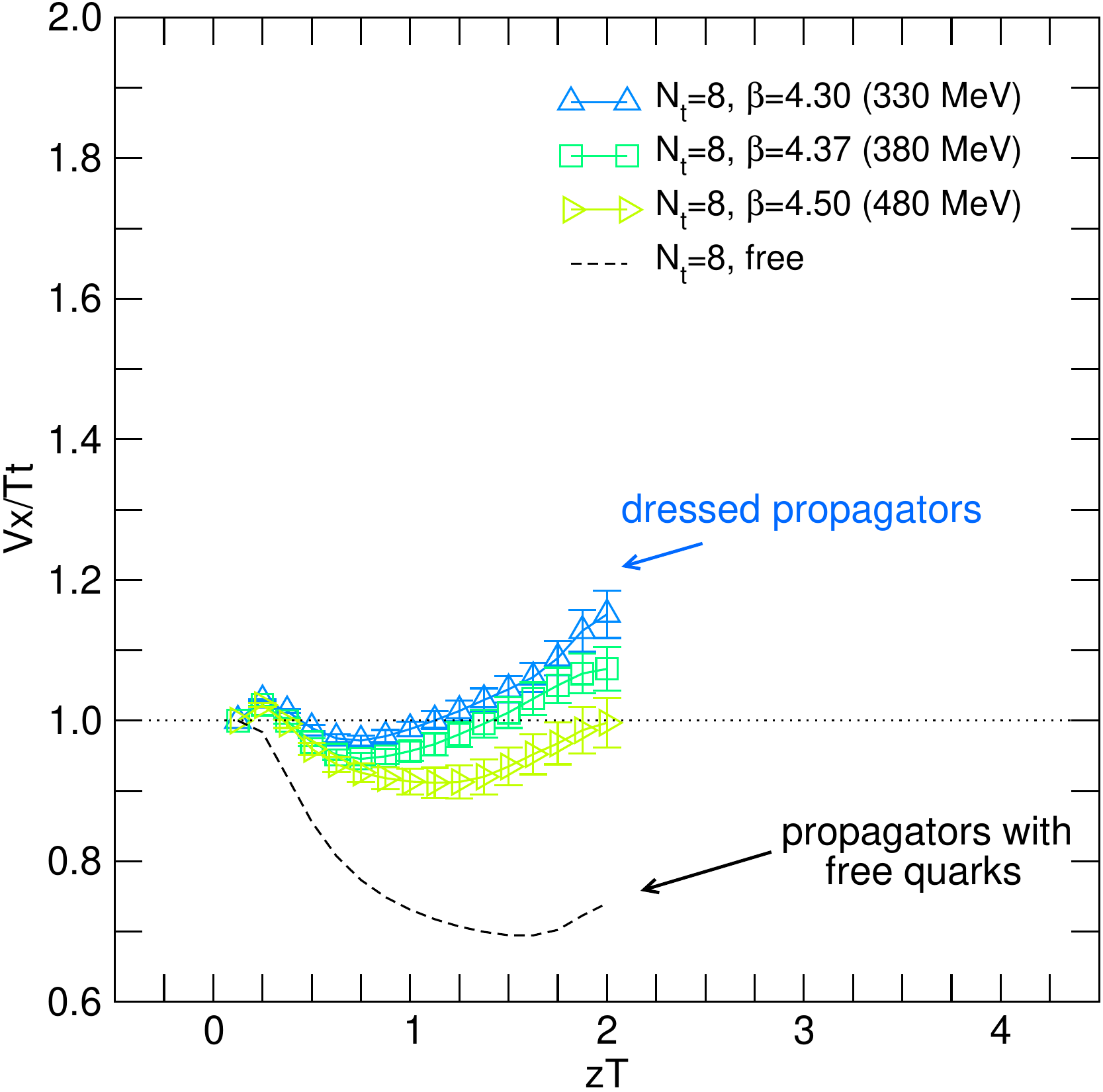} \\
  \includegraphics[scale=0.38]{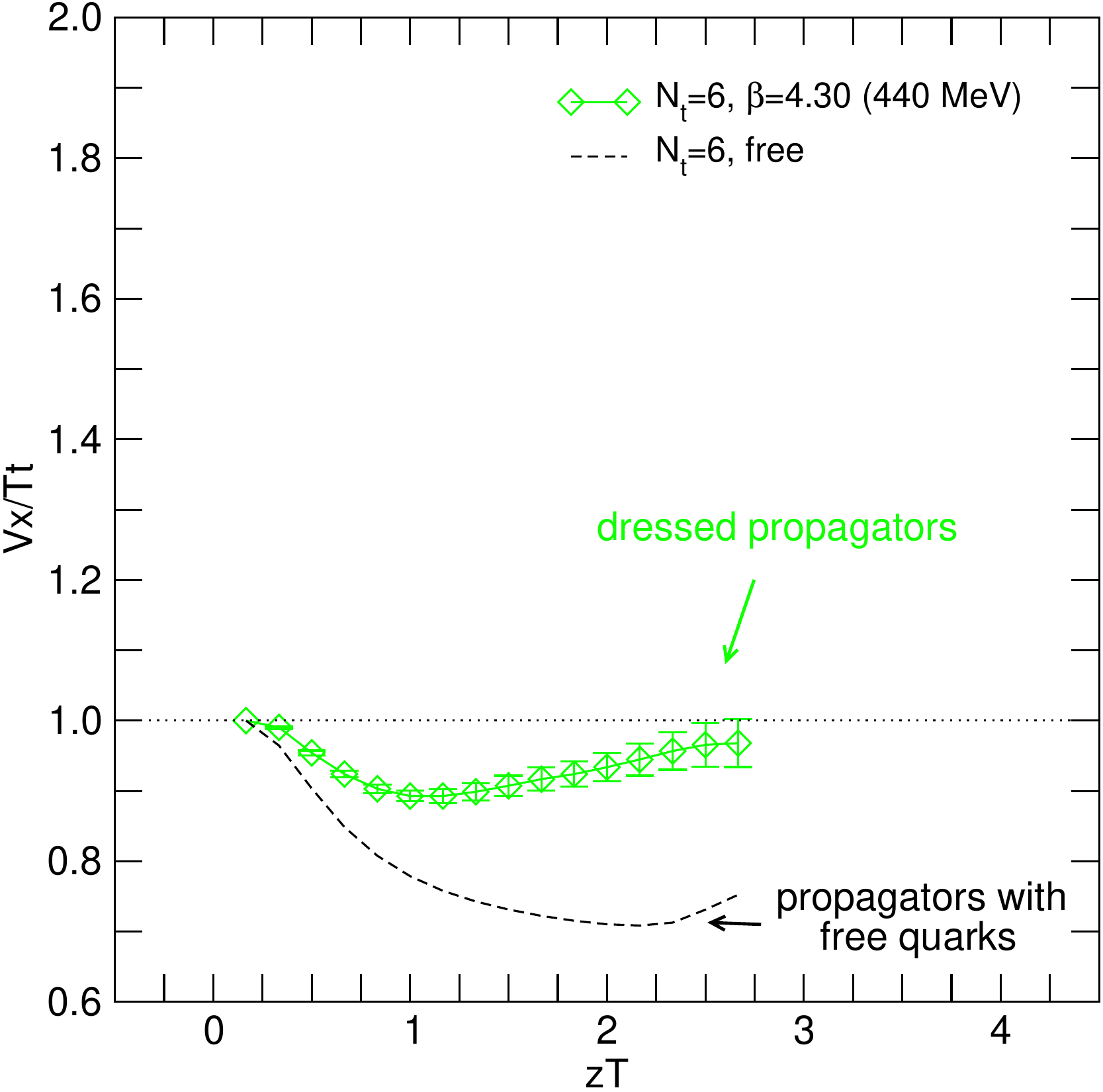}
  \includegraphics[scale=0.38]{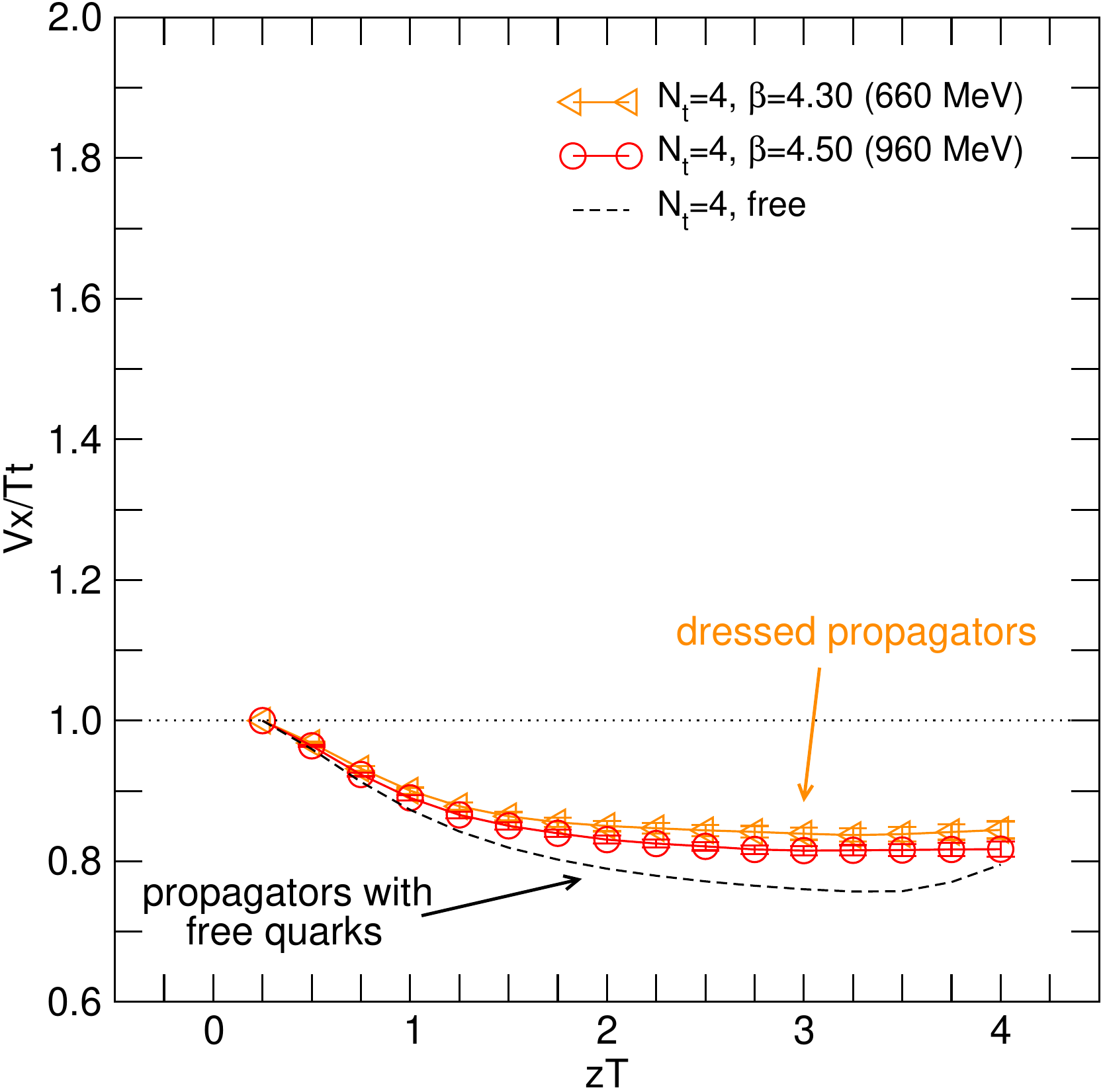}
\caption{
Detailed ratio for correlation functions of operators connected by
$SU(2)_{CS}$. The subplots group lattices of same geometry.
}
\label{allratios}
\end{figure}

The $SU(2)_{CS}$ ratio of correlators for all available
lattice ensembles is shown
in Fig.~\ref{allratios}.
The data therein supports
approximate $SU(2)_{CS}$ symmetry in the
region around $2T_c$. The
interacting correlators
approach the free quark limit at very high
temperatures $T>700$ MeV, where $SU(2)_{CS}$ is broken.
The same ratio for free quarks, \textit{i.e.}
non-interacting quarks, is given for each lattice geometry.
The ratio for free quarks is always below 1, which is in agreement with
the analytic observation that the free (massless) Dirac Lagrangian
$\mathcal{L}= \bar \Psi i \slashed{\partial} \Psi$ breaks $SU(2)_{CS}$ symmetry.

In contrast to free quarks, the situation for interacting quarks is a little more intricate. The
corresponding Lagrangian $\mathcal{L}= \bar \Psi i \slashed{D} \Psi$
in Minkowski space can be written in terms of color-electric and color-magnetic contributions 
\begin{equation}
\mathcal{L}= \bar \Psi i \gamma^0 D_0 \Psi
+ \bar \Psi i \gamma^i D_i \Psi.
\end{equation}
The first term that includes the color-electric interaction in this equation,
is invariant under
$SU(2)_{CS}$ transformations, whereas the second term, with the color-magnetic
contributions,  is not 
\cite{Glozman:2015qva}.
Accordingly, a theory with pure color-electric interaction would show
$SU(2)_{CS}$ symmetry in its spectrum and a $SU(2)_{CS}$ ratio of 1.

Finally, we provide a possible interpretation of our results in Fig.~\ref{allratios}.
For temperatures slightly above the chiral restoration, at 220 MeV, the
$SU(2)_{CS}$ ratio is well above 1 (upper left panel),
which means that the color-magnetic
interaction is large.
Increasing the temperature, its role is diminishing and interaction is
dominantly color-electric. This can be seen by a $SU(2)_{CS}$ ratio of 
approximately 1 at $\sim 2T_c$ (upper right panel of Fig.~\ref{allratios}).
Further increasing temperature, as shown in the lower two panels of
Fig.~\ref{allratios}, the ratio for interacting quarks approaches the limit
for free quarks. Here the dynamics is governed by kinetic contributions, which
are the same as for free quarks. The remaining color-electric interaction slowly
dies out at very high temperature,
and one approaches the asymptotic freedom regime.

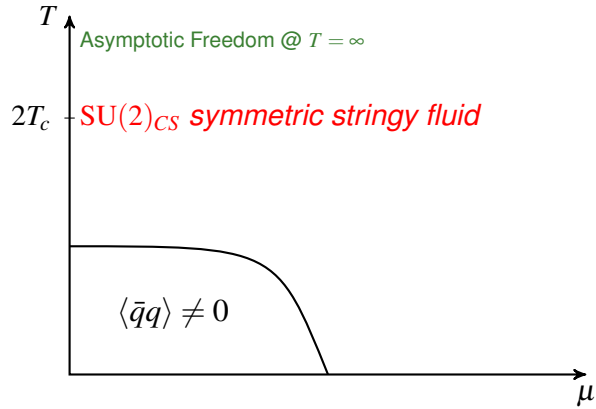
\begin{figure}
  \centering
    \begin{tikzpicture}[scale=0.68]

    \coordinate (y) at (0,7);
    \coordinate (x) at (10,0);
    \coordinate (origin) at (0,0);
    \draw[axis] (y) node[left] {$T$} -- (origin) --  (x) node[below] {$\mu$};
    \draw[-] (0-0.1,5) -- (0+0.1,5) node[left,xshift=-5] {$2T_c$};


    \path
    coordinate (start) at (0,2.5)
    coordinate (c1) at (4,2.5)
    coordinate (end) at (5,0);

    \draw[important line] (start) .. controls (c1) .. (end);

    \node[] at (2,1.2) (hadrons) {\large $\langle \bar q q\rangle \neq 0$};
    \node[right] at (0,6.5) (asymptoticf) {\color{OliveGreen}\scriptsize Asymptotic Freedom @ $T=\infty$};
    \node[right] at (0,5) (su2) {\color{red}\textit{$\mathrm{SU}(2)_{CS}$ symmetric stringy fluid}};
     
  \end{tikzpicture}
\caption{
Sketch of the QCD phase diagram: at $T \sim 2T_c$ the spectrum shows $SU(2)_{CS}$
symmetry. Non-vanishing chemical potential has no effect on this property.
}
\label{pd}
\end{figure}

This might indicate
that elementary objects in QCD at $T \sim 2T_c$ are not free deconfined
quarks but
rather quarks with a definite chirality connected 
by the chromo-electric field \cite{Glozman:2014mka}. 
This conclusion remains also true in matter with
finite chemical potential
\cite{Glozman:2017dfd}, see Fig.~\ref{pd}.

\newpage
\section*{Acknowledgments}
We thank C. B. Lang for numerous discussions.
Support from the Austrian Science Fund (FWF) through the grants
DK W1203-N16 and P26627-N27 is acknowledged.
Numerical calculations are performed on Blue Gene/Q at KEK under its 
Large Scale Simulation Program (No. 16/17-14),
at the Vienna Scientific Cluster (VSC)
and at the cluster of the University of Graz. This work is supported in 
part by JSPS KAKENHI Grant Number JP26247043 and by the Post-K 
supercomputer project through the Joint Institute for Computational 
Fundamental Science (JICFuS).
S.P. acknowledges support from ARRS (J1-8137, P1-0035) and DFG (SFB/TRR 55).


\begin{thebibliography}{99}
\bibitem{Lang:2011vw}
  C.~B.~Lang and M.~Schrock,
  Phys.\ Rev.\ D {\bf 84} (2011) 087704
  doi:10.1103/PhysRevD.84.087704
  [arXiv:1107.5195 [hep-lat]].
\bibitem{Denissenya:2014poa} 
  M.~Denissenya, L.~Y.~Glozman and C.~B.~Lang,
  Phys.\ Rev.\ D {\bf 89}, no. 7, 077502 (2014)
  doi:10.1103/PhysRevD.89.077502
  [arXiv:1402.1887 [hep-lat]].
\bibitem{Denissenya:2014ywa} 
  M.~Denissenya, L.~Y.~Glozman and C.~B.~Lang,
  Phys.\ Rev.\ D {\bf 91}, no. 3, 034505 (2015)
  doi:10.1103/PhysRevD.91.034505
  [arXiv:1410.8751 [hep-lat]].
\bibitem{Denissenya:2015mqa} 
  M.~Denissenya, L.~Y.~Glozman and M.~Pak,
  Phys.\ Rev.\ D {\bf 91}, no. 11, 114512 (2015)
  doi:10.1103/PhysRevD.91.114512
  [arXiv:1505.03285 [hep-lat]].
\bibitem{Denissenya:2015woa} 
  M.~Denissenya, L.~Y.~Glozman and M.~Pak,
  Phys.\ Rev.\ D {\bf 92}, no. 7, 074508 (2015)
  Erratum: [Phys.\ Rev.\ D {\bf 92}, no. 9, 099902 (2015)]
  doi:10.1103/PhysRevD.92.099902, 10.1103/PhysRevD.92.074508
  [arXiv:1508.01413 [hep-lat]].
\bibitem{Glozman:2014mka} 
  L.~Y.~Glozman,
  Eur.\ Phys.\ J.\ A {\bf 51}, no. 3, 27 (2015)
  doi:10.1140/epja/i2015-15027-x
  [arXiv:1407.2798 [hep-ph]].
\bibitem{Glozman:2015qva}
  L.~Y.~Glozman and M.~Pak,
  Phys.\ Rev.\ D {\bf 92} (2015) no.1,  016001
  doi:10.1103/PhysRevD.92.016001
  [arXiv:1504.02323 [hep-lat]].
\bibitem{Lang:2018vuu}
  C.~B.~Lang,
  Phys.\ Rev.\ D {\bf 97} (2018) no.11,  114510
  doi:10.1103/PhysRevD.97.114510
  [arXiv:1803.08693 [hep-ph]].
\bibitem{Rohrhofer:2017grg}
  C.~Rohrhofer, Y.~Aoki, G.~Cossu, H.~Fukaya, L.~Y.~Glozman, S.~Hashimoto, C.~B.~Lang and S.~Prelovsek,
  Phys.\ Rev.\ D {\bf 96} (2017) no.9,  094501
  doi:10.1103/PhysRevD.96.094501
  [arXiv:1707.01881 [hep-lat]].
\bibitem{Brower:2005qw}
  R.~C.~Brower, H.~Neff and K.~Orginos,
  Nucl.\ Phys.\ Proc.\ Suppl.\  {\bf 153} (2006) 191
  doi:10.1016/j.nuclphysbps.2006.01.047
  [hep-lat/0511031].
\bibitem{Brower:2012vk}
  R.~C.~Brower, H.~Neff and K.~Orginos,
  arXiv:1206.5214 [hep-lat].
\bibitem{Tomiya:2016jwr}
  A.~Tomiya, G.~Cossu, S.~Aoki, H.~Fukaya, S.~Hashimoto, T.~Kaneko and J.~Noaki,
  arXiv:1612.01908 [hep-lat].
\bibitem{Cossu:2015kfa}
  G.~Cossu {\it et al.} [JLQCD Collaboration],
  Phys.\ Rev.\ D {\bf 93} (2016) no.3,  034507
  doi:10.1103/PhysRevD.93.034507
  [arXiv:1510.07395 [hep-lat]].
\bibitem{Glozman:2017dfd}
  L.~Y.~Glozman,
  Eur.\ Phys.\ J.\ A {\bf 54} (2018) no.7,  117
  doi:10.1140/epja/i2018-12560-0
  [arXiv:1712.05168 [hep-ph]].
\end{thebibliography}
\end{document}